\documentclass[final,conference]{IEEEtran}
\usepackage{amsmath,amssymb}
\usepackage{mathrsfs}
\usepackage{bm}
\usepackage{graphicx}
\usepackage{epstopdf}
\usepackage{tikz}
\usetikzlibrary{arrows}
\usepackage{pgfplots}
\usepackage{graphicx,booktabs,multirow}
\usepackage[lined,boxed,linesnumbered,commentsnumbered, ruled]{algorithm2e}
\definecolor{colorhkust}{RGB}{20,43,140}
\definecolor{colortsinghua}{RGB}{116,52,129}
\definecolor{color1}{HTML}{D0B22B}

\usetikzlibrary{external}

\newcommand{\bs}{\boldsymbol}

\IEEEoverridecommandlockouts
\begin{document}

\title{Low-Rank Matrix Completion for Mobile Edge Caching in Fog-RAN via Riemannian Optimization
}
\author{\IEEEauthorblockN{Kai Yang$^{*}$, Yuanming Shi$^{*}$, and Zhi Ding$^{\dag}$,
\emph{Fellow, IEEE}}\\

\IEEEauthorblockA{$^{*}$School of Information Science and Technology,
ShanghaiTech University, Shanghai, China\\
$^{\dag}$Dept. of ECE, University of California, Davis, California 95616, USA\\
                           E-mail: \{yangkai, shiym\}@shanghaitech.edu.cn, zding@ucdavis.edu\\
 }
 \thanks{This work is supported by Shanghai Sailing Program No. 16YF1407700.}
}

\maketitle
\IEEEpeerreviewmaketitle
\begin{abstract}
The upcoming big data era is likely to demand tremendous computation and storage resources for communications. By pushing computation and storage to network edges, fog radio access networks (Fog-RAN) can effectively increase network throughput and reduce transmission latency. Furthermore, we can exploit the benefits of cache enabled architecture in Fog-RAN to deliver contents with low latency. Radio access units (RAUs) need content delivery from fog servers through wireline links whereas multiple mobile devices acquire contents from RAUs wirelessly. This work proposes a unified low-rank matrix completion (LRMC) approach to solving the content delivery problem in both wireline and wireless parts of Fog-RAN. To attain a low caching latency, we present a high precision approach with Riemannian trust-region method to solve the challenging LRMC problem by exploiting the quotient manifold geometry of fixed-rank matrices. Numerical results show that the new approach has a faster convergence rate, is able to achieve optimal results, and outperforms other state-of-art algorithms.

\end{abstract}


\IEEEpeerreviewmaketitle

\section{{Introduction}}
The astounding growth of smart mobile device popularity, coupled with new types of wireless services and applications, such as Internet of Things and mobile Cyber-Physical applications, has helped usher in the era of wireless big data \cite{Ding_CMagBigData}. Futuristic high speed networks such as Tactile Internet \cite{Fettweis_JSAC2016} will enable numerous new services and allow for new experiences, e.g., autonomous driving, healthcare, and virtual reality. However, the rapidly growing data and diversified services will put great strain on both storage and computation \cite{shi2015large} in wireless communication systems. This calls for a paradigm shift  for wireless access technologies to provide services with the stringent requirements of ultra-low
latency, high data rate, as well as massive connectivity \cite{Jeff_JSAC5G}.   

By leveraging mobile edge/fog computing concepts \cite{Bonomi2014fog}, the recent proposal of fog radio access network (Fog-RAN) \cite{Yuanming_WCMLargeCVX,peng2016fog} represents a disruptive wireless network architecture to accommodate the upcoming diversified services with ultra-low latency and high data rates. This is achieved by pushing the computation and storage resources to network edges, thereby enabling cloud computation within wireless networks as shown in Fig.~\ref{fig:Fog-RAN}. 

Specifically, in Fog-RAN, each radio access unit (RAU) hosts storage and computation entities, thereby pushing data and its processing closer to end users. This cache enabled network architecture promises to reduce network congestion for wireline communication scenarios \cite{Niesen_TIT2014Caching} and enhance interference coordination in wireless communication networks {\cite{Lau_TSP15Cache}}. By further pushing the content and computation resources to smart mobile devices (MDs), end-to-end latency and network capacity can be improved, e.g., in cache-aided wireless device-to-device (D2D) networks \cite{Caire_TIT2016D2Dcaching}. In particular, for wearable computing applications, offloading computation-intensive applications to proximal smart mobile devices or RAUs can significantly improve roundtrip delay and energy efficiency \cite{Barbarossa_SPM2014MCP}.

In this paper, we mainly focus on the content-centric communication and edge storage aspects of Fog-RAN. The fundamental goal is to leverage the network caching capability for efficient content delivery, thereby reducing end-to-end network latency. To understand this problem, we derive unified interference alignment conditions for content delivery for wireline links between fog server and RAUs, and also for wireless links between RAUs and smart mobile devices. Surprisingly, caching problem with fixed side information is equivalent to the index coding \cite{Jafar_TIT2013TIM} problem, which is equivalent to many other important problems, including topological interference management (TIM) and network coding \cite{Langberg_TIT2015}. Unfortunately, index coding is an NP-hard problem and only a few special cases can been solved efficiently.

In this paper, we present a low-rank matrix completion (LRMC) approach \cite{Candes_2009exactMC} to minimize message delivery latency by exploiting side information provided by the caching capability. The low-rank modeling framework has wide applications in machine learning, computational big data analytics, high-dimensional statistics. Furthermore, the LRMC approach has recently been exploited to solve the wireless topological interference management problem \cite{Yuanming_2016LRMCTWC} and index coding problem \cite{Hassibi_2014matrix}. However, these existing works mainly focus on unicast, in which each message is desired by exactly one destination, whereas in the caching problem each message may be desired by multiple destinations. Despite of the NP-hardness of the resulting LRMC problem, we will reveal that the modeling framework provides algorithmic opportunities.

Although our problem falls within the category of low-rank matrix completion, which has stimulated a flurry of recent research activities \cite{Candes_2009exactMC,Wotao_2012solvingLR, Jain_2013lowAltmin, Vandereycken_ICML2014RiemanMatrixRec}, most prior works are not applicable in our LRMC problem because of their special affine constraint and measurement graph structures. Specifically, the well-known convex relaxation approach by replacing the rank function to the nuclear norm \cite{Candes_2009exactMC} will always yield a full-rank solution in the unicast case, which turns out to be invalid here. Furthermore, alternating minimization approaches \cite{Wotao_2012solvingLR, Jain_2013lowAltmin} also exhibit their own limitations such as slow convergence rate, and impractical assumptions (e.g., incoherence of the original matrix), and available (estimated) ground-truth rank. One exception is the conjugate gradient based Riemannian pursuit algorithm \cite{Vandereycken_ICML2014RiemanMatrixRec}, which explores the embedded manifold structures of fixed-rank matrices to design rank estimation strategies. However, the first-order method has slow convergence and is sensitive to initial points.

In this paper, we develop a Riemannian
trust-region optimization framework \cite{Absil_2009optimizationonManifolds} to provide high accuracy solutions and faster convergence rate for finding the minimal rank in the LRMC problem. This second-order algorithm is more robust to initial point choices, and can obtain reliable solutions. We first solve the fixed-rank matrices, by exploiting the quotient manifold geometry of the search space of fixed-rank matrices \cite{Absil_2009optimizationonManifolds}. We then develop an efficient rank increase mechanism to estimate the minimum rank while satisfying the affine constraint. This is achieved by exploiting the closure of fixed-rank matrices. Numerical results will show that our proposed algorithm can recover existing optimal DoF results in~\cite{Jafar_TIT2013TIM} and outperform state-of-art algorithms.

\section{System Model and Problem Formulation}
\begin{figure}[t]
    \centering
    \includegraphics[width=0.9\columnwidth]{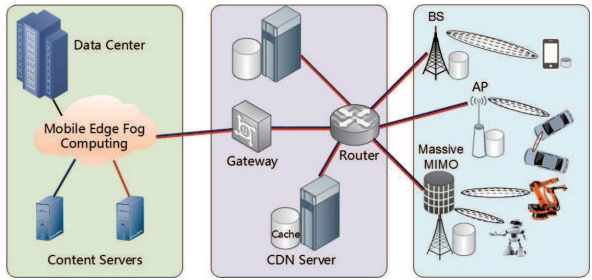}
    \caption{The network architecture of Fog-RAN, in which, the data and its processing are pushed to the edge of networks, including RAUs and smart devices. The benefits of caching allow for wireline communication links between the fog server and RAUs (e.g., massive MIMO, macro base station (BS), and femtocell access point (AP)), as well as the wireless communication links between RAUs and small mobile devices.}
    \label{fig:Fog-RAN}
\end{figure}
Consider a cache-aided Fog-RAN as shown in Fig.~\ref{fig:Fog-RAN}, where both RAUs and smart devices have caching capabilities. To evaluate the benefits of caching in RAUs, we consider the wireline communication network between one fog data center and multiple RAUs. To further investigate the advantages of caching in smart devices, we consider a $K$-user interference channel with cache enabled smart devices. We shall present that the message delivery problems are equivalent to the index coding problem with fixed side information \cite{Jafar_TIT2014indexcoding}.

\subsection{The Benefits of Caching for Wireline Communications} 
\label{wiredcaching}
Consider a wireline communication network between the fog data center and cache-enabled RAUs as shown in Fig. \ref{fig:caching_in_fogran}(a). We assume that the file library consists of $K$ messages $\{W_1,\dots, W_K\}$, each with entropy of $F$ bits. Let $\mathcal{M}=\{1,\dots, K\}$
be the index set of the $K$ independent messages. Define the set $\mathcal{V}_k\subset\mathcal{M}$
as the index set of the messages cached at the $k$-th RAU. We consider the general multiple groupcast scenario, where each message may
be desired by multiple destination nodes \cite{Jafar_TIT2014indexcoding}. Let $\mathcal{M}_k\subseteq\mathcal{M}$ be the index set
of messages desired by RAU $k$. We assume that the available messages will not be further desired
by the corresponding destination node, i.e., $\mathcal{M}_k\cap\mathcal{V}_k=\emptyset$.
In this paper, we are interested in constructing the vector linear coding schemes over real field to maximize the message delivery rate. 

Specifically, the transmitted symbol sequence ${\bm{z}}\in\mathbb{R}^N$ over $N$ channel
uses is given by
\setlength\arraycolsep{1.5pt}
\begin{equation}
{\bm{z}}=\sum_{m\in\mathcal{M}}{\bm{V}}_m{\bm{s}}_m.
\end{equation}
Here, message $W_m$ is split into $Q_m$ scalar data streams, denoted as ${\bm{s}}_m=[s_i(W_m)]_{i=1}^{Q_m}\in\mathbb{R}^{Q_m\times1}$,
and each of which carries one symbol from $\mathbb{R}$ and is transmitted
along the column vectors of the precoding matrix ${\bm{V}}_m\in\mathbb{R}^{N\times
Q_m}$. 

Let ${\bm{U}}_{i,k}\in\mathbb{R}^{Q_i\times N}$ with $ i\in\mathcal{M}_k$
be the receiver combining matrix at RAU $k$ for the desired message
$W_i$. We have the following decoding operation for message $W_i$ at RAU $k$:
\begin{equation}
\hat{\bm{s}}_i=\left({\bm{U}}_{i,k}\bm{V}_i\right)^{-1}\bm{U}_{i,k}\left({\bm{z}}-\sum_{j\in\mathcal{V}_k}{\bm{V}}_j{\bm{s}}_j\right), \forall i\in\mathcal{M}_k.
\end{equation}
To accomplish the above decoding, we impose the following interference alignment conditions \cite{Jafar_TIT2014indexcoding,Yuanming_2016LRMCTWC}:
\begin{eqnarray}
\label{con1}
{\bm{U}}_{i,k}{\bm{V}}_j&=&{\bm{0}}, \forall i\ne j, i\in\mathcal{M}_k,
j\notin {\mathcal{V}}_k\\
\label{con2}
\det({\bm{U}}_{i,k}{\bm{V}}_i)&\ne& 0, \forall i\in\mathcal{M}_k.
\end{eqnarray} 
The first condition (\ref{con1}) is to align and cancel the interference, and the second condition (\ref{con2}) is to preserve the desired signal.  

Therefore, if (\ref{con1}) (\ref{con2}) can both be satisfied
over $N$ channel uses, the message delivery rate tuple $\mathcal{R}=(\frac{Q_1}{N},\frac{Q_2}{N},\cdots,\frac{Q_K}{N})$
can be achieved. In particular, if $Q_1=\cdots=Q_K=Q_0$, we say
that the symmetric data rate is $\frac{Q_0}{N}$. Thus the smaller $N$ we can achieve, the lower latency we can realize. Note that, with $|\mathcal{M}_i|=1$ and $\mathcal{M}_i\cap\mathcal{M}_j=\emptyset, \forall i, j$, we include the multiple unicast scenario as a special case \cite{Yuanming_2016LRMCTWC}. We also observe that the caching problem with fixed side information (i.e., cached message) is equivalent to the index coding problem \cite{Niesen_TIT2014Caching}.      

\subsection{The Benefits of Caching for Wireless Communications}
Recently, caching in content-centric wireless networks is a very popular topic \cite{wang2016mobility,tao2015content}. We consider a wireless communication modeled as the $K$-user interference channel as shown in Fig. 2 (b). Assume that each transmitter $k$ (i.e., RAU) has the message $W_k$ for transmission. Let the set $\mathcal{V}_k\subset\mathcal{M}$
as the index set of the messages cached at the $k$-th receiver (i.e., smart device). Let $\mathcal{M}_k\subseteq\mathcal{M}$ with $\mathcal{M}_k\cap\mathcal{V}_k=\emptyset$ be the index set
of messages desired by receiver $k$. With message splitting as in Section {\ref{wiredcaching}}, over the $N$ channel uses, the received signal $\bm{y}_i\in\mathbb{C}^N$ at  receiver $k$ is given by
\begin{equation}
\label{inout1}
{\bm{y}}_k=\sum_{
i=1}^K{\bm{H}}^{[ki]}{\bm{V}}_i{\bm{s}}_{i}+{\bm{n}}_k,
\forall k,
\end{equation} 
where ${\bm{n}}_k\sim\mathcal{CN}({\bm{0}}, {\bm{I}}_N)$ is the additive isotropic
white Gaussian noise, ${\bm{s}}_i\in\mathbb{C}^{Q_i\times1}$ represents message $W_i$, ${\bm{H}}^{[ki]}={H}_{ki}{\bm{I}}_N$
is an $N\times N$ diagonal
matrix with  $H_{ki}\in\mathbb{C}$ as the  constant channel
coefficient between transmitter $i$ and receiver $k$ over $N$ channel uses in the considered block. 

Let ${\bm{U}}_{i,k}\in\mathbb{C}^{Q_i\times N}$ with $ i\in\mathcal{M}_k$
be the receiver combining matrix at receiver $k$ for the desired message
$W_i$. We assume that the channel state information $H_{ki}$ with $i\in\mathcal{V}_k$ is available at receiver $k$. Therefore, we can first  eliminate the undesired messages available in $\mathcal{V}_k$, resulting the interference space as $\sum_{j\notin\mathcal{V}_k, i\neq j}{\bm{H}}^{[kj]}{\bm{V}}_j$ for the message $W_i$. In the regime of asymptotically high SNR, to accomplish decoding, we impose
the constraint that the desired signal space ${\bm{H}}^{[ki]}{\bm{V}}_i$ is complementary to the interference space, resulting the following interference alignment condition (see \cite{Jafar_TIT2013TIM,Yuanming_2016LRMCTWC})
\begin{eqnarray}
\label{con1s}
{\bm{U}}_{i,k}\bm{H}^{[kj]}{\bm{V}}_j&=&{\bm{0}}, \forall i\ne j, i\in\mathcal{M}_k,
j\notin {\mathcal{V}}_k\\
\label{con2s}
\det({\bm{U}}_{i,k}\bm{H}^{[ki]}{\bm{V}}_i)&\ne& 0, \forall i\in\mathcal{M}_k.
\end{eqnarray}
Since ${\bm{H}}^{[ki]}=H_{ki}{\bf{I}}_N$ for the constant
channel over the $N$ channel uses, conditions (\ref{con1s}) and (\ref{con2s}) are equivalent to conditions (\ref{con1}) and (\ref{con2}), respectively.     

Therefore, if (\ref{con1s}) (\ref{con2s}) are both satisfied over $N$ channel uses, the parallel
interference-free channels can be obtained. Therefore, the DoF tuples $\mathcal{D}=(\frac{Q_1}{N},\frac{Q_2}{N},\cdots,\frac{Q_K}{N})$ for the message delivery over wireless interference channel.  In particular, if $Q_1=\cdots=Q_K=Q_0$, we say that the the symmetric DoF $\frac{Q_0}{N}$ can be achieved for each message. As DoF provides a first-order characterization for capacity, we shall minimize channel uses $N$ to improve the message delivery data rate.

\section{Low-Rank Matrix Completion Approach for Mobile Edge Caching}
In this section, we present a low-rank matrix completion modeling framework
to find the minimal number of channel uses such that the interference alignment
conditions (\ref{con1}) and
(\ref{con2}) can be satisfied. Specifically, let $\bm{X}_{ij,k}={\bm{U}}_{i,k}{\bm{V}}_j$. Without lose of generality, we set ${\bm{U}}_{i,k}{\bm{V}}_i=\bm{I}_{Q_i\times Q_i}, \forall i\in\mathcal{M}_k$. We thus can restrict $\bm{X}_{ij,k}$'s to the real field without losing any performance in terms of achievable DoFs \cite{Yuanming_2016LRMCTWC}. Therefore, for $\bm{X}_{ij,k}\in\mathbb{R}^{Q_i\times Q_j}$, we have
\begin{eqnarray}
\label{IAC}
{\bm{X}}_{ij,k} = \left\{ \begin{array}{ll}
 \bm{0}&~~ \forall i\ne j, i\in\mathcal{M}_k,
j\notin {\mathcal{V}}_k\\
 \bm{I} &~~  \forall i=j, i\in\mathcal{M}_k\\
 \star &~~ \textrm{otherwise},
  \end{array} \right.
\end{eqnarray}
where $\star$ represents  arbitrary  $Q_i\times Q_j$ vectors.

Let $\bm{X}=[\bm{X}_{ij,k}]\in\mathbb{R}^{M\times Q}$ with $M=\sum_{i=1}^{K}|\mathcal{M}_i|$ and $Q=\sum_{i=1}^KQ_i$. Based on (\ref{IAC}), the interference alignment conditions (\ref{con1}) and (\ref{con2}) can be rewritten as
\begin{eqnarray}
\mathcal{P}_{\Omega}(\bm{X})=\bm{J},
\end{eqnarray} 
where $\mathcal{P}_{\Omega}:\mathbb{R}^{M\times Q}\rightarrow\mathbb{R}^{M\times Q}$ is the orthogonal projection operator onto the subspace of matrices which vanish outside $\Omega$ such that the $(i,j)$-th component of $\mathcal{P}_{\Omega}(\bm{X})$ equals to $X_{ij}$ if $(i,j)\in\Omega$ and zero otherwise. Here, the constant $M\times Q$ matrix $\bm{J}=[\bm{J}_{mj}]$ is given by
\begin{eqnarray}
\label{IAC1}
{\bm{J}}_{mj} = \left\{ \begin{aligned}
 \bm{I}&~~  {\textrm{if}}~\sum_{i=1}^{k-1}|\mathcal{M}_i|+1\le m\le\sum_{i=1}^{k}|\mathcal{M}_i|, j\in\mathcal{M}_k\\
 \bm{0} &~~ \textrm{otherwise},
  \end{aligned} \right.~~
\end{eqnarray}
where $\bm{I}$ is the $Q_j\times Q_j$ identity matrix. The set $\Omega$ is constructed as $\Omega=\{\mathcal{S}_m\times\mathcal{S}_j|\sum_{i=1}^{k-1}|\mathcal{M}_i|+1\le m\le\sum_{i=1}^{k}|\mathcal{M}_i|,~{\textrm{with}}~j\notin\mathcal{V}_k~\textrm{or}~j\in\mathcal{M}_k\}$, where $\mathcal{S}_i=\{1+\sum_{k=1}^{i-1}Q_k,\dots,\sum_{k=1}^i Q_k\}$.

\subsection{Low-Rank Matrix Completion Modeling Framework}
Observe that the rank of matrix $\bm{X}\in\mathbb{R}^{M\times Q}$ equals $N$, we propose to solve the following LRMC problem
\begin{eqnarray}\label{problem:LRMC}
\mathop{\textrm{minimize }}_{\bm{X}}&& \textrm{rank}(\bm{X}) \nonumber\\
\textrm{subject to}&& \mathcal{P}_{\Omega}(\bm{X})=\bm{J}
\end{eqnarray}
to maximize the message delivery rate while satisfying interference alignment conditions (\ref{con1}) (\ref{con2}), thereby reducing latency.

\begin{figure}[h]
    \centering
    \includegraphics[width=0.8\columnwidth]{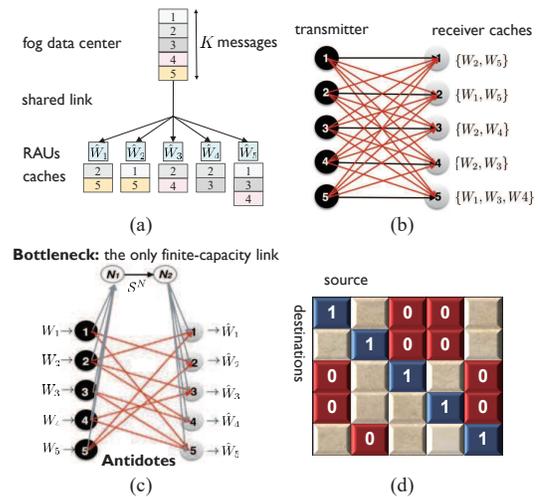}
    \caption{Cache enabled Fog-RAN. (a) A wireline communication network between fog data center and cache-enabled RAUs. (b) Wireless cache-enabled networks between RAUs and cache-enabled smart devices. (c) The corresponding index coding problem. Red lines represent the side information. For example, receiver$~1$ has $\mathcal{V}_1=\{W_2,W_5\}$ as side information. (d) Associated low-rank matrix completion model. In this example no message splitting is taken into consideration, therefore $Q_i=1$. Gray cells represent elements to be arbitrary values. For example, $(3,1)$-th element of $\bm{X}$ must be $0$ since $W_1$ is not in the side information of receiver$~3$.}
    \label{fig:caching_in_fogran}
\end{figure}

An illustrative example of a caching problem with $5$ messages and $3$ destination
nodes is given in Fig.~\ref{fig:caching_in_fogran}. 
In this setting, we have
$\mathcal{V}_1=\{2,5\}$, $\mathcal{V}_2=\{1,5\}$, $\mathcal{V}_3=\{2,4\}$, $\mathcal{V}_4=\{2,3\}$, $\mathcal{V}_5=\{1,3,4\}$. Then all elements of the set $\Omega$ are $\{(1,1),(1,3),(1,4),(2,2),(2,3),(2,4)$, $(3,1),(3,3),(3,5),(4,1),(4,4),(4,5),(5,2),(5,5)\}$.

\subsection{Problem Analysis}
Although LRMC problem is NP-hard generally, there are several popular methods
to yield a feasible solution. The most common approach is to relax the
rank function to nuclear norm $\bm{X}_*$ \cite{Candes_2009exactMC}. But obviously
in the unicast case, because $\bm{X}_*\geq|\textrm{Tr}(\bm{X})|=Q$ always
holds, we would always end up with a full rank solution. Another common way to solve
the LRMC problem applies \textit{alternating minimization}. In this
algorithm, under a fixed rank $r$ we can factorize $\bm{X}$ as $\bm{X}=\bm{U}\bm{V}^T$. Next we optimize $\bm{U}\in\mathbb{R}^{M\times r}$ and $\bm{V}\in\mathbb{R}^{Q\times
r}$ alternatively. Unfortunately, this method has a very slow convergence rate and
the rank assumption is impractical since the rank is unknown for our problem.
Additionally, the algorithm calls for the incoherence property of matrix
$\bm{X}$, which is also invalid. Interestingly, the authors of \cite{Vandereycken_ICML2014RiemanMatrixRec} presented a Riemannian pursuit algorithm based on conjugate gradient method, termed as EmbG. EmbG takes a rank increase strategy by exploiting
the embedded manifold structures of fixed rank matrices. As a first-order
algorithm, the convergence of EmbG is slow, and is sensitive to initial
points. Therefore, we hope to find a method with higher accuracy, faster convergence rate and more robust to initial points.

\section{Riemannian Trust-Region Method for Low-Rank Matrix Completion on Quotient Manifolds}

In this section, we shall present a Riemannian trust-region framework which is a second-order algorithm for faster convergence rate and higher resolution. We first recast the LRMC problem as a set of fixed-rank subproblems and present the corresponding Riemannian trust region approach. Then we develop our rank-increasing strategy by exploiting the closure of fixed-rank matrices to ensure that the objective function monotonically decreases.

\subsection{Fixed-Rank Riemannian Trust-Region for Matrix Completion}
By defining the cost function with Frobenius norm
\begin{eqnarray}
        f(\bm{X})=\frac{1}{2}\|\mathcal{P}_{\Omega}(\bm{X})-{\bm{J}}\|_F^2,
\end{eqnarray}
problem (\ref{problem:LRMC}) can be reformulated as a sequence of subproblems $\mathscr{P}$ given the fixed rank $r$:
\begin{eqnarray}
        \mathscr{P}:\quad \mathop{\textrm{minimize}}_{\bm{X}\in\mathcal{M}_r} \frac{1}{2}\|\mathcal{P}_{\Omega}(\bm{X})-{\bm{J}}\|_F^2.
\end{eqnarray}
With a little abuse of notation here, $\mathcal{M}_r:=\{\bm{X}\in\mathbb{R}^{M\times Q}:~\textrm{rank}(\bm{X})=r\} $ is a smooth ($C^{\infty}$) manifold.

The matrix $\bm{X}$ can be represented by a SVD-type factorization of well defined manifold $\bm{X}=\bm{U}\bm{\Sigma}\bm{V}^T,~\bm{U}\in\textrm{St}(r,M),\bm{\Sigma}\in\textrm{GL}(r),\bm{V}\in\textrm{St}(r,Q)$. Here $\textrm{St}(r,M)=\{\bm{Y}\in\mathbb{R}^{M\times r}:\bm{Y}^T\bm{Y}=\bm{I}_r\} $ is the \textit{compact Stiefel manifold} of orthonormal $M\times r$ matrices, and $\textrm{GL}(r)=\{\bm{Y}\in\mathbb{R}^{r\times r}: \textrm{rank}(\bm{Y})=r\} $ is the \textit{non-compact Stiefel manifold} of all $r\times r$ nonsingular matrices. Thus $\bm{X}=(\bm{U},\bm{\Sigma},\bm{V})$ and \textit{computational space} is $\mathcal{M}_r:=\textrm{St}(r,M)\times\textrm{GL}(r)\times\textrm{St}(r,Q)$, on which our algorithm is built. Then the \textit{quotient space} is $\mathcal{M}_r/\sim:=\mathcal{M}_r/(\mathcal{O}(r)\times\mathcal{O}(r))$, $\mathcal{O}(r)=\{\bm{Y}\in\mathbb{R}^{r\times r}: \bm{Y}^T\bm{Y}=\bm{I}_r\} $ denote the $r$-order \textit{orthogonal group}.

Our algorithm is based on trust region method on Riemannian manifold. The trust-region subproblem in the quotient manifold $\mathcal{M}_r\slash\sim$ is horizontally lifted to the \textit{horizontal space} $\mathcal{H}_{\bm{X}}$, and formulated as
\begin{eqnarray}\label{problem:sub trust-region}
\mathop{\rm{minimize}}_{{\bm{\xi}}_{\bm{X}}\in\mathcal{H}_{\bm{X}} \mathcal{M}_{r}}&& {m({\bm{\xi}}_{\bm{X}})}\nonumber\\
{\rm{subject~to}}&& g_{\bm{X}}({\bm{\xi}}_{\bm{X}},{\bm{\xi}}_{\bm{X}})\leq \delta^2.
\end{eqnarray}
The trust-region radius is $\delta$ and the cost function is given by
\begin{eqnarray}
        m({\bm{\xi}}_{\bm{X}}) =&& f(\bm{X})+g_{\bm{X}}({\bm{\xi}}_{\bm{X}},\textrm{grad}_{\bm{X}}f)+ \nonumber\\
         &&\frac{1}{2} g_{\bm{X}}({\bm{\xi}}_{\bm{X}},\textrm{Hess}_{\bs{X}}f\left[\bm{\xi}_{\bm{X}}\right]).
\end{eqnarray}

To develop an effective algorithm, we need to determine
\begin{itemize}
    \item Riemannian metric: $g_{\bm{X}}$
    \item Riemannian gradient: $\textrm{grad}_{\bm{X}}f$
    \item Riemannian Hessian: $\textrm{Hess}_{\bm{X}}f[\bm{\xi}_{\bm{X}}]$
\end{itemize}
$\textrm{grad}_{\bm{X}}f$ and $\textrm{Hess}_{\bs{X}}f$ are the generalization of Euclidean gradient and Hessian to the Riemannian manifold~\cite{Absil_2009optimizationonManifolds}.

\textit{1) Riemannian metric:} The Riemannian metric must have identical matrix representation along the equivalent class $[\bm{X}]$. By taking second-order derivative of $f(\bm{X})$, namely $\bm{\mathcal{L}}_{\bm{X}\bm{X}}(\bm{X})$, we can induce its following Riemannian metric from the approximation \cite{Yuanming_2016LRMCTWC}
\begin{eqnarray}
        g_{\bm{X}}(\bm{\xi}_{\bm{X}},\bm{\zeta}_{\bm{X}})\approx&\langle \bm{\xi}_{U},\bm{\zeta}_{U}\bm{\Sigma}\bm{\Sigma}^T\rangle + \langle \bm{\xi}_{\Sigma},\bm{\zeta}_{\Sigma}\rangle \nonumber\\
        &+ \langle \bm{\xi}_{V},\bm{\zeta}_{V}\bm{\Sigma}^T\bm{\Sigma}\rangle,
\end{eqnarray}
in which $\bm{\xi}_{\bm{X}}=(\bm{\xi}_{U},\bm{\xi}_{\Sigma},\bm{\xi}_{V}),\bm{\zeta}_{\bm{X}}=(\bm{\zeta}_{U},\bm{\zeta}_{\Sigma},\bm{\zeta}_{V})\in T_{\bm{X}}\mathcal{M}_r $ and $\bm{X}\in(\bm{U},\bm{\Sigma},\bm{V})$.

\textit{2) Riemannian gradient:} Euclidean gradient is $\nabla f(\bm{X})=\mathcal{P}_{\Omega}(\bm{X})-{\bm{J}}$. Riemannian gradient $\textrm{grad}f(\bm{X})$ gives an orthogonal projection of $\nabla f(\bm{X})$ to the tangent space, i.e.
\begin{gather}\label{eq:grad}
        \textrm{grad}f(\bm{X})=\Pi_{T_{\bm{X}}\mathcal{M}_r}(\nabla f(\bm{X})) \\
        \Pi_{T_{\bm{X}}\mathcal{M}_r}(\bm{G}): \bm{G}\rightarrowtail P_{U}\bm{G}P_V+{P_{U}^\perp}\bm{G}P_V+{P_{U}}\bm{G}{P_V^\perp},
\end{gather}
in which $P_{U}=\bm{U}\bm{U}^T$ and $P_{U}^{\perp}=\bm{I}-\bm{U}\bm{U}^T$ for $\bm{X}=\bm{U}\textrm{diag}(\bm{\sigma})\bm{V}^T$ and $\bm{\sigma}\in\mathbb{R}^r$.

\textit{3) Riemannian Hessian:} Much like Riemannian gradient, Riemannian Hessian has the form of
\begin{equation}\label{eq:hessian}
        \textrm{Hess}_{\bm{X}}f[\bm{\xi}_{\bm{X}}]=\Pi_{\mathcal{H}_{\bm{X}}\mathcal{M}_r}(\nabla_{\bm{\xi}_{\bm{X}}} \textrm{grad}f(\bm{X})).
\end{equation}
Let us define the \textit{Riemannian connection} $\nabla_{\bm{\eta}_{\bm{X}}}\bm{\xi}_{\bm{X}}$, and substitute the Riemannian gradient and Riemannian connection into (\ref{eq:hessian}). From the \textit{Koszul formula}, its connection to Euclidean directional derivative $D\bm{\xi}_{\bm{X}}[\bm{\eta}_{\bm{X}}]$
\begin{equation}
        \nabla_{\bm{\eta}_{\bm{X}}}\bm{\xi}_{\bm{X}}=D\bm{\xi}_{\bm{X}}[\bm{\eta}_{\bm{X}}]+(\bm{\theta}_{U},\bm{\theta}_{\Sigma},\bm{\theta}_{V}),
\end{equation}
where $D\bm{\xi}_{\bm{X}}$ denotes the Euclidean directional derivative, $\bm{\theta}=(\bm{\theta}_U,\bm{\theta}_{\Sigma},\bm{\theta}_V)$ is related to the solutions of Lyapunov equations.

With Riemannian metric, Riemannian gradient and Riemannian Hessian defined, a fixed-rank Riemannian trust-region algorithm has been well developed and can be implemented in \textit{Manopt} package~\cite{boumal2014manopt}.

\subsection{Rank Increase Algorithm}\label{subsection:rank-increasing}
To solve the rank minimization problem (\ref{problem:LRMC}) with the fixed rank trust-region algorithm, we develop a rank-increasing strategy based on the closure of the set of fixed-rank matrices, $\mathcal{M}_{\leq r}=\{\bm{X}\in\mathbb{R}^{M\times Q}: \textrm{rank}(\bm{X})\leq r\}$.

\SetNlSty{textbf}{}{:}
\IncMargin{1em}
\begin{algorithm}[ht]
\textbf{Input:} $\Omega,{\bm{J}}$, accuracy $\epsilon$.\\
 Initialize: $\bm{X}_0^{[1]}\in\mathbb{R}^{m\times n}$, $(m,n)=\textrm{size}({\bm{J}})$.\\
 \While{rank-deficient and not converged}{
  Compute a solution to the trust-region subproblem $\mathscr{P}$ with initial point $\bm{X}_0^{[r]}$ for fixed rank $r$ with Riemannian optimization.\\
   Compute the value of cost function $f(\bm{X})$.\\
   Update rank $r$ to $r+1$ with rank increase strategy (\ref{eq:rankincreasestrategy}) in Section~\ref{subsection:rank-increasing}.
 }
 \textbf{Output:} $\bm{X}^{[r]}$ and rank $r$.
 \caption{Riemannian Trust-Region Algorithm for LRMC Problem~(\ref{problem:LRMC})}
 \label{algorithm:rankpursuit}
\end{algorithm}
\begin{table*}[!t]
    \renewcommand{\arraystretch}{1.3}
    \caption{Recover the Optimal DoF Results in Cases with/without Message Splitting in~\cite{Jafar_TIT2013TIM} via LRMC}
    \label{tab:achievableDoF1}
    \centering
    \begin{tabular}{c||c|c|c|c||c|c}
    Network Topology in~\cite{Jafar_TIT2013TIM} & Fig. 4 & Fig. 6 & Fig. 12(a) & Fig. 12(b) & Fig. 9(a) & Fig. 10(a)\\
    \hline
    DoF in~\cite{Jafar_TIT2013TIM} & 1/2 & 1/2 & 1/4 & 1/3 & 3/7 & 2/5 \\
    \hline
    Number of Data Streams ($Q_i$) & 1 & 1 & 1 & 1 & 3 & 2 \\
    \hline
    Optimal Rank by LRMC Approach ($n$) & 2 & 2 & 4 & 3 & 7 & 5
    \end{tabular}
\end{table*}

Consider the $(r+1)$-th step of iteration where we compute $\bm{X}_{r+1}$ from $\bm{X}_r$.
To escape from $\mathcal{M}_r$ and embed $\bm{X}_r$ to $\mathcal{M}_{r+1}$, we can give a good initial point by using the linear-search method
\begin{equation}
        \bm{X}_{r+1}=\mathcal{R}_{{r+1}}(\bm{X}_r+\alpha_r\bm{\Xi}_r),
\end{equation}
where we choose the negative Euclidean gradient in the \textit{tangent cone} $T_{\bm{X}_r}\mathcal{M}_{r+1}$ as our search direction $\bm{\Xi}_r$  at $\bm{X}_r$ \cite{schneider2015convergence} with step-size $\alpha_r$ and $\mathcal{R}_{r+1}$ denotes the retraction to computation space $\mathcal{M}_{r+1}$. Then $\bm{\Xi}=\arg\min_{\bm{\Xi}\in T_{\bm{X}_r}\mathcal{M}_{\leq r+1}}\|-\nabla_{\bm{X}_r}f-\bm{\Xi}\|_F=-\textrm{grad}_{\bm{X}_r}f+\bm{\Xi}_r^{(1)}$. $\bm{\Xi}_r^{(1)}$ is the orthogonal projection on the tangent space $T_{\bm{X}_r}\mathcal{M}_r$. Hence, our rank updating strategy can be formulated as
\begin{equation}\label{eq:rankincreasestrategy}
        \bm{X}_{r+1}=\mathcal{R}_{r+1}(\bm{X}_r+\alpha_r(\bm{\Xi}_r^{(1)}-\textrm{grad}_{\bm{X}_r}f)),
\end{equation}
which keeps the cost function $f$ decreasing monotonically.

Thus far, through solving fixed-rank subproblem and rank increase strategy, a complete Riemannian trust-region algorithm can be used to solve the caching problem in Fog-RAN.

\section{Numerical Results}
In this section we first test the convergence rate of our trust-region approach. We then show that our algorithm can achieve optimality in cases that \cite{Jafar_TIT2014indexcoding} refers to. Finally, we run the whole simulation on $20$ messages and destination nodes with $3$ data streams for varying cache size varies from no caching to just one message needs to be transmitted. The following two known algorithms are compared:
\begin{itemize}
        \item EmbG: This algorithm \cite{Vandereycken_ICML2014RiemanMatrixRec} is developed on the embedded manifold via fixed-rank optimization \cite{vandereycken2013low} with the Riemannian pursuit rank increase strategy \cite{Vandereycken_ICML2014RiemanMatrixRec}.
        \item LMaFit: This algorithm introduces the alternating minimization scheme to solve problem $\mathscr{P}$ \cite{Wotao_2012solvingLR}.
\end{itemize}

\subsection{Convergence Rate}
\begin{figure}[ht]
        \centering
        \includegraphics[width=0.85\columnwidth]{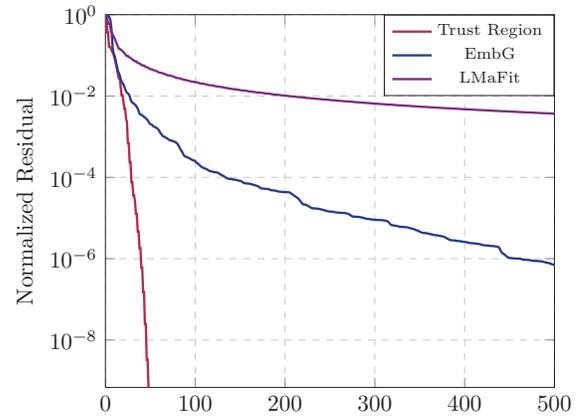}
        \caption{Convergence rate when fixing the rank of $\bm{X}$ as $40$.}
        \label{fig:convergence}
\end{figure}

Consider a caching problem where the cache size is $10$ and each message is exactly desired by one destination node, the number of messages and destination nodes are both $30$ and each message is split to $5$ data streams. We investigate the convergence rate of all three algorithms given a fixed rank $r=40$. Fig.~\ref{fig:convergence} shows that the trust-region method outperforms the two competing schemes in convergence rate and can achieve a higher precision solution as expected. The convergence rate of LMaFit is lower than other two manifold approach.
\subsection{Achieve the Existing Optimal DoF Results}\label{section:5b}

In this part of experiments, we consider equivalent model for topological interference management  in~\cite{Jafar_TIT2013TIM} where optimal DoF is given. To check whether or not the Riemannian trust-region method achieves the optimal DoF, both of the unicast cases with and without message splitting are tested. Table~\ref{tab:achievableDoF1} shows that the LRMC algorithm can numerically achieve all of the optimal DoF results in~\cite{Jafar_TIT2013TIM}.

\subsection{Data Rates for Different Cache Size}
\begin{figure}[!t]
        \centering
        \includegraphics[width=0.85\columnwidth]{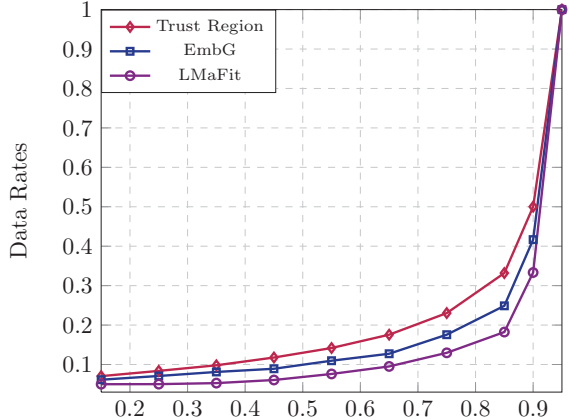}
        \caption{Average data rates varying from cache sizes. Each message has 3 data streams. The cache size is normalized and tolerance of the cost is $10^{-7}$.}
        \label{fig:result caching}
\end{figure}

In order to evaluate the performance of the Riemannian trust-region method in caching networks with different cache sizes, we test a simple unicast model with 20 messages and users with 3 data streams for each message. The data rate we can achieve is $3/n$ where $n$ is the solution of algorithms. The cache size $m$ varies from $0$ to $19$ since $m=20$ represents that no message is needed. The average data rates results are shown in Fig.~\ref{fig:result caching}. In our tests, the side information is generated randomly and 50 experiments for each cache size are performed. When the cost falls below $\epsilon=10^{-7}$, rank increase strategy stops.

From the experimental result, we observe that Riemannian trust-region approach can achieve a higher data rates than two other algorithms. And EmbG outperforms alternating projection method (LMaFit).

\section{Conclusions}
In this paper, we propose a unified low-rank matrix completion approach for content delivering of caching problem in Fog-RAN. We connected the caching problem with LRMC problem, and presented a Riemannian trust-region algorithms, solving the fixed-rank subproblem and giving an rank increase strategy that can guarantee monotonic decrease of objective function. Numerical results show that our approach can achieve optimal solution in existing cases for which the optimal value is known and outperforms algorithms such as EmbG and LMaFit. One interesting future work is the determination of conditions under which the Riemannian trust-region algorithm can guarantee optimality.

\bibliographystyle{ieeetr}

\end{document}